\documentclass[prb,nofootinbib,twocolumn]{revtex4} 

\usepackage{graphicx}
\usepackage{dcolumn}
\usepackage{bm}
\usepackage{threeparttable}
\usepackage{times}
\usepackage{mathptmx}
\usepackage{lscape}
\usepackage{natbib}
\usepackage{amsmath}
\usepackage{amssymb}
\usepackage{braket}
\usepackage{comment}
\usepackage{color}

\def\degree{\kern-.2em\r{}\kern-.3em}

\begin{document}

\title{ Representation-Aware Transport-Information Measure for Non-Inclusive Discrete Supports  }

\author{Koretaka Yuge$^{\dagger}$}
\affiliation{
Department of Materials Science and Engineering,  Kyoto University, Sakyo, Kyoto 606-8501, Japan\\
}%

\begin{abstract}
{ 
Information-theoretic measures for comparing probability distributions are widely used across physics and other fields. When two discrete distributions have non-inclusive supports, however, the Kullback-Leibler (KL) divergence is in general not directly applicable, and various alternative divergences and distances have been introduced. These measures compare the resulting distributions themselves, but do not generally retain information about the representation transformations by which the discrete distributions are generated from underlying continuous ones.
Here we introduce a ``representation-aware transport-information measure'' for discrete distributions with non-inclusive supports, formulated based on the standard KL divergence. We consider two continuous reference distributions, each transformed into a discrete representation through its own discretization scheme. Rather than comparing only the resulting discrete distributions or their continuous references, we additionally retain local information associated with the representation-change schemes. The resulting measure can therefore distinguish discrete representations that may have identical discrete probability landscapes but originate from different continuous references or discretization schemes.
The construction is based on the transport-information cost of continuous-to-discrete representation in the framework of unavoidable canonical nonlinearity (UCN). UCN provides a non-arbitrary correspondence between the transport cost of discretization as an extrinsic geometric operation and the information-theoretic indistinguishability of nearby continuous distributions, thereby allowing a discrete representation to be associated with a local family of underlying continuous distributions on the statistical manifold. As an illustrative application, we consider the canonical nonlinearity of classical configurational systems with different crystal structures, whose discrete supports need not be mutually inclusive. Their continuous Gaussian reference distributions provide representation-independent baselines, while their distinct discretization schemes retain information associated with the underlying lattice structures. The proposed measure enables the information-theoretic comparison of such discrete representations.

}
\end{abstract}

\maketitle

\section{Introduction}
Comparing probability distributions is a fundamental task in various fields including statistical physics, information theory, and data analysis. For discrete probability distributions, the Kullback-Leibler (KL) divergence\cite{KL} can provide a natural measure of their information difference when the support of the reference distribution is identical or inclusive. In many practical situations, however, discrete distributions are typically defined on different configuration spaces, discretization grids, or sets of realizable states, so that their supports are not mutually inclusive. In such cases, the KL divergence between the discrete distributions can be ill-defined, even when both distributions originate from well-defined underlying continuous distributions. This situation is particularly natural when the discrete distributions are obtained through different representation schemes from their continuous ones: the difference in their supports is then not merely a statistical difference between two probability distributions, but also reflects the transformations used to represent the underlying systems.
 
Several measures have been developed to compare discrete probability distributions without requiring such a support-inclusion relation. Examples include the Jensen-Shannon divergence,\cite{JS} Hellinger distance,\cite{HL} Wasserstein distance,\cite{WS} and maximum mean discrepancy.\cite{MM} These measures provide well-defined comparisons for discrete distributions with different supports, and are useful when the objective is to characterize the difference in landscape between the probability distributions themselves. However, when discrete distributions arise from distinct representation transformations of underlying continuous distributions, another question naturally arises: whether the difference associated with the representations can be evaluated together with the difference between the resulting distributions. A measure based only on the final discrete distributions cannot, in general, distinguish two representations that happen to produce the same discrete probability distribution, nor can it retain information about the set of continuous distributions and discretization schemes from which those discrete representations originated. Thus, in addition to comparing probability distributions as statistical objects alone, it is useful to consider a comparison of discrete representations together with the transformations that generate them from their continuous references.  

This perspective is closely related to the transport-information cost associated with continuous-to-discrete representation change developed in our previous work on unavoidable canonical nonlinearity (UCN).\cite{UCN} In that framework, a continuous probability distribution is transformed into a discrete counterpart through a prescribed local discretization scheme, and the resulting discretization  cost is characterized from the local geometry of the continuous statistical family together with the geometry of the discretization cells: In the UCN, the discretization cost is expressed by the standard KL divergence expected over appropriate parameter variations on statistical manifold, without introducing any arbitrary structure. 
Thus, the UCN formulation can provide a natural way to retain information about the transformation from a continuous to its discrete counterpart, rather than treating the resulting discrete distribution as an isolated statistical object. The present work extends this idea from the self-comparison of a distribution before and after discretization, to the comparison of two discrete distributions generated from potentially different continuous distributions and different discretization schemes. In particular, the two discrete supports are not required to be mutually inclusive. The proposed measure for discrete distributions on inclusive or non-inclusive supports is given by the standard KL divergence between their \textit{continous} reference distributions, integrated over parameter variations associated with their individual discretization geometry: the measure certainly distinguish dicrete distributions with their underlying continuous references as well as their discretization scheme. The details are shown below.

\section{Concept and Construction}
\subsubsection*{Representation-Aware Comparison}
The central idea of the present construction is to regard a discrete representation not only as a static probability distribution on a discrete support, but as the result of a transformation from an underlying continuous probability distribution. We therefore distinguish between the continuous reference and the discretization scheme used to obtain the discrete distribution. A discrete representation is characterized schematically as
\begin{eqnarray}
(P_c,\mathcal D)\longrightarrow P_d,
\end{eqnarray}
where $P_c$ denotes the continuous reference distribution and $\mathcal D$ denotes the discretization scheme, including the local structure of the discretization cells. The resulting discrete distribution $P_d$ is then the observable output of this transformation.

This distinction becomes important when two representations are compared. Consider two representations
\begin{eqnarray}
(P_{c1},\mathcal D_1)\longrightarrow P_1, \qquad (P_{c2},\mathcal D_2)\longrightarrow P_2,
\end{eqnarray}
where the continuous reference distributions and the discretization schemes may both differ. The supports of $P_1$ and $P_2$ are not assumed to be mutually inclusive. More importantly, the pair $\left(P_1,P_2\right)$ alone does not necessarily contain all the information associated with their two representations. For example, two different continuous distributions may give rise to the same discrete distribution under an appropriate discretization, while the underlying continuous distributions remain different. Conversely, the same continuous distribution may be represented by different discretization schemes, producing different discrete representations. These cases are indistinguishable if only the final discrete probability distributions are considered.

We therefore seek a measure that compares the two representations while retaining the information associated with the underlying continuous distributions and their discretization schemes. The key point is that the discretization itself induces an \textit{ensemble} of local neighborhood of continuous distributions around each reference distribution. Rather than assigning an arbitrary distance directly between the discrete supports, we compare these underlying continuous distributions and average the comparison over the neighborhoods associated with the two discretization schemes.

\subsubsection*{Construction of the Measure}
Let us consider the two continuous reference distributions $P_{c1}$ and $P_{c2}$, where they are individually parameterized by $\xi$ and $\eta$ on their statistical submanifold, namely,
\begin{eqnarray}
P_{c1}=P_{c1:\xi},
\qquad
P_{c2}=P_{c2:\eta}.
\end{eqnarray}
We also consider their corresponding discrete counter part $P_{1}$ and $P_{2}$, which is explicitly constructed through the specific discretization scheme discussed later.
Note that since the discretization itself is an \textit{extrinsic} geometric operation on sample space living outside the information geometry, the choice of the parameters $\xi$ and $\eta$ here is confined to those identifiable with support coordinates $q$ satisfying\cite{UCN}
\begin{eqnarray}
\label{eq:GL}
^{\forall}A \in \textrm{GL}\left(f\right): \quad q \mapsto A q \quad \Rightarrow \quad \xi \mapsto A \xi,
\end{eqnarray}
and parameter $\eta$ also satisfies the same condition of $\eta \mapsto A \eta$. The corresponding Fisher metrics for $\xi$ and $\eta$ are given by $\Omega^{\left( 1 \right)}$ and $\Omega^{\left( 2 \right)}$.

We here also focus on the specific discretization procedure called ``local-map (LM) scheme'' based on the translation of any bounded convex set $\omega\in \mathbb{R}^{f}$ as discretization cell $V$ to fullfill the sample space $\simeq \mathbb{R}^{f}$, and continuous probability mass within the cell is mapped onto its center $q'$. This can be concretely represented for the $k$-th cell $V_{k}$ by 
\begin{eqnarray}
\label{eq:disc}
^{\forall} k, \,\, P_{d}\left( q'_{k} \right) = \int_{V_{k}} P_{c}\left( q \right) dq.
\end{eqnarray}
Therefore, the discretization scheme naturally corresponds to performing coarse-graining of the continuous distributions, which locally maps its probability mass onto underlying discretization cells.  

Under these preparations, when we consinder the LM-scheme discritization of continuous distribution $P_{c:\xi}$, we have shown that at vanishing discretization scale limit $V_{k}\to 0$, the following correspondence always holds on:\cite{UCN}
\begin{eqnarray}
\label{eq:ucn}
\widetilde{W}_{2}^{2} \left( P_{c:\xi}, P_{d} \right) = 2 \mathbb{E}_{\rho\left( \delta\xi \right)} \left[ D\left( P_{c:\xi+\delta\xi}:P_{c:\xi} \right) \right] = \textrm{Tr}\left( M\Omega \right),
\end{eqnarray}
where  $\widetilde{W}_{2}$ denotes 2-Wasserstein-based transport cost, where the transport plan is not optimized but employing the LM scheme of Eq.~\eqref{eq:disc}, $D$ the KL divergence, $\Omega$ the Fisher metric for $P_{c:\xi}$, and $\rho\left( \delta\xi \right)$ denotes probabily distribution for parameter variation $\delta\xi$, where $\delta\xi$ is extended to $f$-dimensional  random variable. The distribution 
$\rho\left( \delta\xi \right)$ is constructed so that its covariance matrix is proportional to the second-moment matrix $M$ for the discretization cell $\omega$, defined as 
\begin{eqnarray}
M = \dfrac{1}{V_{\omega} } \int_{\omega} u u^{T} du,
\end{eqnarray}
where the coordinate $u$ is chosen such that its mean over $\omega$ vanishes. 
Eq.~\eqref{eq:ucn} certainly exhibits that the squared discretization cost on sample space, $\widetilde{W}_{2}^{2} \left( P_{c:\xi}, P_{d} \right)$ is identical to the cumulative information-geometric indistuinguishability around $P_{c}$, $D\left( P_{c:\xi+\delta\xi}:P_{c:\xi}\right)$, averaged over its parameter variations $\delta\xi$ generated by discretization geometry through $\textrm{cov}\left[\rho\right]\propto M$. 
The important point in Eq.~\eqref{eq:ucn} is that the distribution $\rho$ characterizes the local uncertainty in the continuous representation associated with the LM discretization schemes. 
This suggests that, when a discrete distribution $P_d$ is represented on the continuous statistical manifold, it should be associated not with a single point $P_{c:\xi}$, but with a local family $\left\{P_{c:\xi+\delta\xi}\right\}$ whose parameter fluctuations are determined by the discretization geometry.

Therefore, we naturally introduce an UCN-based conceptual identification of the discrete distribution as
\begin{eqnarray}
\label{eq:pcpd}
P_{d} \sim \left\{ P_{c:\xi+\delta\xi} | \textrm{cov}\left[ \rho\left( \delta\xi \right) \right] = M \right\}.
\end{eqnarray}
This definition, especially of $\textrm{cov}\left[ \rho\left( \delta\xi \right) \right] = M$ by itself, does not specify a unique statistical submanifold for the parameter $\xi$ satisfying the condition of Eq.~\eqref{eq:GL}.  To obtain a definite representation-aware measure, a common parameterization must therefore be fixed.
In the present construction, we fix the submanifold to the translation family of the continuous reference distribution, namely, 
\begin{eqnarray}
P_{c:\xi}\left(x\right)=P_c\left(x-\xi\right),
\end{eqnarray}
and identify its location coordinate $\xi$ with the corresponding support coordinate $x$. Hence, $\delta\xi=\delta x$, 
so that the covariance of the local parameter fluctuations is directly identified with the second-moment matrix of the discretization cell, i.e., $\textrm{cov}[\rho\left(\delta\xi\right)]=M$, leading to the definition of Eq.~\eqref{eq:pcpd}. Note that generally, selected submanifold with parameter $\xi$ satisfying Eq.~\eqref{eq:GL} does not always results in $\textrm{cov}[\rho\left(\delta\xi\right)]=M$, yet the proportional relation $\textrm{cov}[\rho\left(\delta\xi\right)]\propto M$ still retains. 
This specific choice of a local family fixes a common statistical submanifold and the relative parameter scale entering the following transport-information measure of $D_{\textrm{RA}}$, so that their values obtained from different continuous reference distributions and different LM-based discretization schemes can be compared on the same information scale. More general choices of statistical submanifold would be considered as extensions, but are not part of the present definition of $D_{\textrm{RA}}$.

Under this identification, we can now introduce representation-aware transport-information measure between discrete distributions $P_{1}$ and $P_{2}$, defined as 
\begin{widetext}
\begin{eqnarray}
\label{eq:DRA}
D'_{\mathrm{RA}}(P_1:P_2) = \int d\left( \delta\xi \right) d\left( \delta\eta \right) \rho_{1}\left( \delta\xi \right)\rho_{2}\left( \delta\eta \right) D\left( P_{c1:\xi+\delta\xi}: P_{c2:\eta+\delta\eta} \right). 
\end{eqnarray}
\end{widetext}
The proposed measure does not therefore compare $P_1$ and $P_2$ directly on their discrete statistical manifold. Instead, it compares the continuous reference distributions underlying the two representations, while averaging over their local neighborhoods induced by their respective discretization geometry through $\rho_{1}$ and $\rho_{2}$. 

An important property of Eq.~\eqref{eq:DRA} is that it does not require either discrete support to contain another one. The KL divergence appearing in Eq.~\eqref{eq:DRA} is evaluated between the continuous reference distributions rather than between the discrete distributions. Thus, the construction remains well defined even when the two discrete representations have non-inclusive or disjoint supports, provided that the corresponding continuous KL divergences are well defined.

\subsubsection*{Local Information Available from the Discretization}
The construction in Eq.~\eqref{eq:DRA} appears requiring the full information about distributions $\rho_{1}$ and $\rho_{2}$ of the local parameter displacements, \textit{unless} there provides any prior information about the parameter variations. However, in the UCN framework as discussed, the LM scheme is introduced at vanishing discretization scale limit, and the required information is confined to (i) their parameter variations for the first moments, 
\begin{eqnarray}
\int 	d\left( \delta\xi  \right)\left[ \rho_{1}\left( \delta\xi \right) \delta\xi \right] = 0 \nonumber \\
\int 	d\left( \delta\eta \right) \left[ \rho_{2}\left( \delta\eta \right) \delta\eta \right] = 0
\end{eqnarray}
and (ii) the second-moments
\begin{eqnarray}
M_1
&=&
\int d(\delta\xi)\,\rho_1(\delta\xi)\,\delta\xi\,\delta\xi^{\mathsf T}, \nonumber \\
M_2
&=&
\int d(\delta\eta)\,\rho_2(\delta\eta)\,\delta\eta\,\delta\eta^{\mathsf T}.
\end{eqnarray}
These conditions naturally arises from the UCN construction that the discretization scheme is characterized locally by the second moment of the corresponding discretization cell for transport cost as well as by the second moment of the parameter variations, $\rho_{1}$ and $\rho_{2}$.\cite{UCN} Note that the restriction to the second moments is not merely a choice of a particular discretization limit. In the local UCN construction, the information associated with the discretization is obtained from the infinitesimal transport-information correspondence. Consequently, the leading information that can be associated with a local cell is determined by its second-order geometry together with the Fisher information of the continuous statistical family. Information about higher-order moments of the cell and parameter fluctuations are not essentially required by this local correspondence. 

This point is fundamentally important for the present construction. The quantity in Eq.~\eqref{eq:DRA} with no constraints contains, in principle, information from all orders of the local displacement distributions. Such information would require additional specification of the distributions $\rho_{1}$ and $\rho_{2}$, and would therefore introduce structure beyond that fixed by the UCN construction. 
Within the local UCN correspondence, the discretization information entering the information-geometric side is specified by these second moments. We therefore use $M_1$ and $M_2$ to obtain the concrete form below.

\subsubsection*{Concrete Formulation of $D_{\textrm{RA}}$}
To obtain a concrete form determined by the local information specified in the above discussions, we expand the KL divergence in Eq.~\eqref{eq:DRA} around the two reference distributions. Let us define
\begin{eqnarray}
F(\xi,\eta) = D\left(P_{c1:\xi}:P_{c2:\eta}\right).
\end{eqnarray}
For small displacements $\delta\xi$ and $\delta\eta$, we write $F$ up to the second-order as 
\begin{widetext}
\begin{eqnarray}
F(\xi+\delta\xi,\eta+\delta\eta) =F(\xi,\eta) +\nabla_\xi F\cdot \delta\xi +\nabla_\eta F\cdot \delta\eta +\frac12 \delta\xi^{\mathsf T}A\,\delta\xi +\delta\xi^{\mathsf T}B\,\delta\eta +\frac12 \delta\eta^{\mathsf T}C\,\delta\eta, 
\end{eqnarray}
\end{widetext}
where
\begin{eqnarray}
A_{ij} = \frac{\partial^2F}{\partial\xi_i\partial\xi_j}, \,
B_{i\alpha} = \frac{\partial^2F}{\partial\xi_i\partial\eta_\alpha}, \,
C_{\alpha\beta} = \frac{\partial^2F}{\partial\eta_\alpha\partial\eta_\beta}.
\end{eqnarray}
Because the local displacement distributions are centered, the first-order terms vanish upon averaging. We additionally assume that the local displacements associated with the two discretization schemes are statistically independent. The mixed second-order contribution then also vanishes:
\begin{eqnarray}
&&\int d(\delta\xi)d(\delta\eta)\rho_1(\delta\xi)\rho_2(\delta\eta) \delta\xi^{\textrm{T}}B\delta\eta \nonumber \\
&&= \left[ \int d(\delta\xi)\rho_1(\delta\xi)\delta\xi \right]^{\textrm{T}} B \left[ \int d(\delta\eta)\rho_2(\delta\eta)\delta\eta \right] \nonumber \\
&& =0.
\end{eqnarray}
Using the second moments introduced, we finally obtain the UCN-based representation of Eq.~\eqref{eq:DRA}, given by 
\begin{eqnarray}
\label{eq:raf}
\boxed{
D_{\mathrm{RA}}\left( P_{1}:P_{2} \right) = D\left(P_{c1}:P_{c2}\right) + \frac12\textrm{Tr}\left( M_1A \right) + \frac12\textrm{Tr}\left( M_2C \right)
} \nonumber \\
\,
\end{eqnarray}
Equation~\eqref{eq:raf} provides the UCN-constructible form of the representation-aware measure for discrete distributions. The first term describes the information difference between the two continuous reference distributions as the standard KL divergence, while the remaining terms describe the deformation of this information difference induced by the two discretization schemes. Thus, even when the continuous reference distributions are identical, different discretization schemes can give rise to a nonzero representation-dependent contribution.

It is worth emphasizing that the matrices $A$ and $C$ in Eq.~\eqref{eq:raf} are Hessians of a KL divergence and are not, in general, identical to the Fisher metric of the two continuous statistical families. In particular, the Fisher metric arises naturally in the special local self-comparison in which the two arguments of the KL divergence, $P_{c1}$ and $P_{c2}$ coincide, whereas the present construction allows them to be different. The present measure therefore does not require the two continuous reference distributions to be close or to belong to the same statistical family. 
For example, for the first argument, we obtain
\begin{eqnarray}
A_{ij} = \Omega^{(1)}_{ij} + \mathbb{E}_{P_{c1}} \left[ \frac{\partial^2\log P_{c1}} {\partial\xi_i\partial\xi_j} \log\frac{P_{c1}}{P_{c2}}
\right],
\end{eqnarray}
Similarly, 
\begin{eqnarray}
C_{\alpha\beta} = -\mathbb{E}_{P_{c1}} \left[ \frac{\partial^2\log P_{c2}} {\partial\eta_\alpha\partial\eta_\beta} \right].
\end{eqnarray}
Consequently, neither correction term in Eq.~\eqref{eq:raf} should in general be interpreted as a Fisher-induced information difference.  

Eq.~\eqref{eq:raf} should be distinguished from the measure in Eq.~\eqref{eq:DRA} without any constraint in the parameter variation eometry. The latter is an average of the standard KL divergences and is therefore non-negative. The second-order expression of Eq.~\eqref{eq:raf}, on the other hand, is a local representation determined through the second moments of the discretization schemes. Its discretization-induced contribution, 
\begin{eqnarray}
\Delta_{\textrm{disc}} = \frac12\textrm{Tr}\left( M_1A \right) + \frac12\textrm{Tr}\left( M_2C \right),
\end{eqnarray}
is not required to be non-negative in the general cross-comparison considered here. Accordingly,  $D_{\textrm{RA}}$ is not assumed to be a divergence in the strict mathematical sense. Rather, it is a local representation-aware transport-information measure that separates the difference of the continuous reference distributions from the information deformation associated with their discretization.

Finaly, we note that the same construction can be straightforwardly applied when only one of the two representations is discretized. For example, when the first representation is continuoius and the second is discrete, we obtain 
\begin{eqnarray}
D_{\textrm{RA}}(P_{c1}:P_{2}) = D\left(P_{c1}:P_{c2}\right) + \frac{1}{2}\textrm{Tr}\left(M_2C\right).
\end{eqnarray}
Thus, the representation-induced contribution associated with the discretized representation remains explicitly incorporated, while no such contribution appears for the continuous representation.

\subsubsection*{$D_{\textrm{RA}}$ for Common Continuous Reference}
A particularly transparent case for the proposed measure is obtained when the two discrete representations share the same continuous reference distribution but are generated by different discretization schemes. Namely, consider
\begin{eqnarray}
\left( P_c,\mathcal{D}_1 \right)\longrightarrow P_1, \qquad \left( P_c,\mathcal{D}_2 \right)\longrightarrow P_2,
\end{eqnarray}
where $P_c$ is common to both discrete representations, while the individual discretization geometries are characterized by their second-moment matrices $M_1$ and $M_2$, respectively.
In this case, the continuous-reference contribution to Eq.~\eqref{eq:raf} vanishes identically,
$D\left( P_c:P_c \right)=0$.

Moreover, since the two representations belong to the same continuous statistical family, the Fisher metric $\Omega$ is common to both representations. For the common translation family considered above, the two Hessians appearing in Eq.~\eqref{eq:raf} therefore reduce to the same Fisher metric, i.e., $A=C=\Omega$.
Consequently, the representation-aware measure takes the particularly simple form
\begin{eqnarray}
\label{eq:rafc}
\boxed{
D_{\mathrm{RA}}\left( P_1:P_2 \right)  = \frac{1}{2}\textrm{Tr}\left( M_1\Omega \right) + \frac{1}{2}\textrm{Tr}\left( M_2\Omega \right)
}.
\end{eqnarray}
This expression provides a direct interpretation of $D_{\textrm{RA}}$ in terms of the discretization cost associated with each representation. In fact, according to the UCN correspondence,
\begin{eqnarray}
\textrm{Tr}\left( M_i\Omega \right) = 2\mathbb{E}_{\rho_i} \left[D\left( P_{c:\xi+\delta\xi}:P_{c:\xi} \right)\right]
\end{eqnarray}
for each discretization scheme $\mathcal{D}_i$. Hence, $D_{\textrm{RA}}$ in Eq.~\eqref{eq:rafc} 
can be interpreted as the sum of information cost associated with the two discretization schemes. Thus, even when the underlying continuous information geometry is identical, the two discrete representations can acquire different information-theoretic contributions through their distinct discretization geometries $M_1$ and $M_2$ together with the Fisher metric $\Omega$ for underlying continuous reference.

It is worth emphasizing that $M_i$ characterizes the local second-order geometry of the discretization scheme rather than the discrete support itself. Therefore, equality of the discrete support does not necessarily imply equality of the representation-induced information cost. For example, different partitions can, in principle, have the same set of cell centers while possessing different cell shapes and hence different second-moment matrices $M_i$. Conversely, different discrete supports are generally led from different local discretization geometries. In this sense, $D_{\textrm{RA}}$ retains information about the discretization procedure that is not contained in the final discrete probability distribution alone. 
The above result also clarifies the role of the common continuous reference in the representation-aware measure. When $P_c$ is shared, the distinction between the two discrete representations is not encoded in a difference between continuous statistical manifolds; it is encoded entirely, to second order, in how the common continuous geometry is mapped and represented through $M_1$ and $M_2$. Thus, the representation dependence is transferred from the continuous reference distribution to the local geometry of the discretization schemes.

\subsubsection*{JSD-based representation-aware measure}

The construction of $D_{\textrm RA}$ can also be formulated using a divergence other than the Kullback-Leibler (KL) divergence. This is particularly useful when the continuous reference distributions have non-inclusive supports, for which the KL divergence may be undefined or infinite. As a practical example, we consider the Jensen-Shannon divergence (JSD),\cite{JS} which remains finite for arbitrary probability distributions defined with respect to a common dominating measure, including distributions with non-overlapping supports. The JSD is defined as
\begin{eqnarray}
D_{\rm JS}\left( P:Q \right) &=& \frac{1}{2}D_{\rm KL}\left( P:R \right) +\frac{1}{2}D_{\rm KL}\left( Q:R \right), \nonumber \\
R&=&\frac{P+Q}{2}.
\end{eqnarray}

We define the corresponding representation-aware measure by simply replacing the KL divergence in the local-family construction of 
$D'_{\rm RA}$ with the JSD:
\begin{widetext}
\begin{eqnarray}
\begin{aligned}
D_{\rm RA}^{\prime \rm JS}\left( P_1:P_2 \right) = \int d(\delta\xi)\,d(\delta\eta)\, \rho_1(\delta\xi)\rho_2(\delta\eta) D_{\rm JS}\left(P_{c1:\xi+\delta\xi}: P_{c2:\eta+\delta\eta} \right).
\end{aligned}
\end{eqnarray}
\end{widetext}
As in the KL-based construction, we assume centered, independent parameter fluctuations with second-moment matrices $M_1$ and $M_2$. Defining
\begin{eqnarray}
F_{\rm JS}(\xi,\eta) = D_{\rm JS}\left(P_{c1:\xi}:P_{c2:\eta}\right),
\end{eqnarray}
and its Hessians
\begin{eqnarray}
A_{ij}^{\textrm{JS}} = \frac{\partial^2F_{\textrm{JS}}}{\partial\xi_i\partial\xi_j}, \,\,
C_{\alpha\beta}^{\textrm{JS}} = \frac{\partial^2F_{\textrm{JS}}}{\partial\eta_\alpha\partial\eta_\beta},
\end{eqnarray}
the second-order formulation straightforwardly becomes
\begin{eqnarray}
\boxed{
\begin{aligned}
D_{\rm RA}^{\rm JS}\left(P_1:P_2\right) = D_{\rm JS}\left(P_{c1}:P_{c2}\right) +\frac12\textrm{Tr}(M_1A^{\rm JS}) +\frac12\textrm{Tr}(M_2C^{\rm JS}). 
\end{aligned}
} \nonumber \\
\quad
\end{eqnarray}
Thus, the structure of the resulting measure is essentially unchanged; only the reference divergence and its associated local Hessians differ. The JSD is adopted here as a representative practical choice, rather than as a unique prescription. Other divergences constructed from suitable mixtures of probability distributions may also be considered, depending on the comparison objective and the properties required of the resulting measure.

An instructive special case arises when the two continuous reference distributions coincide and the same translation-family parameterization is used. At a common parameter value, let $P_{c1}=P_{c2}=P_c$, and denote the Fisher information matrix of this family by $\Omega$. For the equally weighted JSD, the second-order structure at coincident distributions is given by
\begin{eqnarray}
D_{\rm JS}\left(P_{c:\xi+\delta\xi}: P_{c:\xi+\delta\eta}\right) = \frac18\left(\delta\xi-\delta\eta\right)^{T} \Omega\left(\delta\xi-\delta\eta\right).  
\end{eqnarray}
Consequently, the two Hessians satisfy
\begin{eqnarray}
A^{\rm JS}=C^{\rm JS}=\frac14\Omega
\end{eqnarray}
at the common reference distribution. Since the reference JSD vanishes, the second-order representation-aware measure reduces to
\begin{eqnarray}
\boxed{
D_{\rm RA}^{\rm JS}\left(P_1:P_2\right) = \frac18\textrm{Tr}(M_1\Omega) +\frac18\textrm{Tr}(M_2\Omega).
}
\end{eqnarray}
This expression clearly shows that even when the continuous reference distributions are identical, the representation-aware contribution can remain nonzero if the two discretization schemes have nonzero second-moment matrices. The measure therefore retains information associated with the discretization schemes that is absent from the divergence between the continuous references alone.

\subsubsection*{Connection to Discrete Configurational Systems}
The motivation for the present construction arises naturally in classical discrete configurational systems. Consider, for example, configuration spaces associated with different crystal structures, or with different compositions within the same crystal structure. When the configurational degrees of freedom are represented by correlation functions, the corresponding discrete configurational supports are generally not mutually inclusive. The difference in the supports is determined by the underlying lattice and composition, and therefore represents a structural difference between the discrete systems rather than merely a difference in the probability weights assigned to a common set of states.

This situation is directly relevant to the canonical nonlinearity of classical discrete systems, which corresponds to the nonlinear character of canonical average from a set of many-body interactions to equilibrium configuration.\cite{CN} For a given discrete configurational system, the hypothetical Gaussian reference provides the globaly linear reference to measure the nonlinerity. 
The Gaussian reference is then mapped onto the discrete configurational support of that system, yielding a discrete Gaussian representation. Thus, when two configurational systems have different supports, their respective continuous Gaussian references $P_{{G}}$s are discretized according to different representation schemes, namely, 
\begin{eqnarray}
(P_{G1},\mathcal D_1)\longrightarrow P_{dG1},
\qquad
(P_{G2},\mathcal D_2)\longrightarrow P_{dG2},
\end{eqnarray}
where$P_{dG1}$ and $P_{dG2}$ denote the corresponding discrete Gaussian representations obtained by the LM scheme.

The comparison between such discrete Gaussian representations is of particular interest because these distributions serve as references for the canonical nonlinearity of the respective discrete systems. A direct comparison of the final discrete distributions, however, does not necessarily retain all of the information relevant to this comparison. For example, two different configurational systems may have different continuous Gaussian references but yield identical probability weights on the common part of their discrete supports. A comparison based only on the resulting discrete probability distributions would then assign zero KL divergence to those common states, despite the fact that the underlying continuous references and the discretization schemes are different.

Conversely, two configurational systems may have identical mean vectors and covariance matrices for their CDOS. Their continuous Gaussian reference distributions are then identical, and a comparison performed only at the continuous level gives no information difference. Nevertheless, mapping this common Gaussian reference onto two different configurational supports generally produces different discrete Gaussian representations. Since these supports are determined by the lattice and composition, the resulting difference reflects the distinct discrete representations of the two configurational systems.

These examples illustrate why the information associated with the continuous-to-discrete transformation is relevant when comparing discrete Gaussian references across different configurational systems. The representation-aware construction introduced above certainly provides a way to retain these information.

\section{Conclusions}

We have introduced a representation-aware transport-information measure for comparing discrete probability distributions with non-inclusive supports. The central idea is to distinguish a discrete distribution from the representation transformation that generates it. Instead of comparing only the resulting discrete distributions, we consider the continuous reference distributions underlying them together with their respective discretization schemes.

The construction is motivated by the transport-information correspondence in the UCN framework, which identifies the transport cost of continuous-to-discrete distribution with the information-theoretic indistinguishability of nearby continuous distributions. This correspondence provides a natural conceptual identification of a discrete representation with a local family of continuous distributions whose parameter fluctuations are determined by the discretization geometry. The proposed measure then compares two such representation-associated families through the standard KL divergence.

Using the local information specified by the UCN framework, we obtained the concrete second-order form of the measure $D_{\textrm{RA}}$, which separates into the standard KL divergence between the two continuous reference distributions and a discretization-induced information deformation. The latter contribution is not required to be non-negative, and $D_{\textrm{RA}}$ is therefore not interpreted as a divergence in the strict mathematical sense.
The resulting construction extends information-theoretic comparison from probability distributions themselves to the representations through which they are generated. It can therefore retain information that is lost when only the final discrete probability distributions are compared. The connection to discrete configurational systems discussed above illustrates how this representation-aware viewpoint can be relevant when comparing configurational spaces associated with different lattice structures or compositions.

\section{Acknowledgement}
This work was supported by Research Grant from Hitachi Metals$\cdot$Materials Science Foundation.

\end{document}